\documentclass[english]{article}
\pdfoutput=1
\usepackage{amsmath}
\usepackage{amssymb}
\usepackage{epsfig,cite}

\makeatletter
\usepackage{babel}

\def\be#1{\begin{equation}\label{#1}}
           
          \def\ee{\end{equation}}

\begin{document}
\begin{titlepage}
\thispagestyle{empty}

\bigskip

\begin{center}
\noindent{\Large \textbf
{Massive $p-$form trapping as a $p-$form on a brane}}\\

\vspace{0,5cm}

\noindent{I. C. Jardim ${}^{a}$\footnote{e-mail: jardim@fisica.ufc.br},G. Alencar  ${}^{a}$\footnote{e-mail: geovamaciel@gmail.com }, R. R. Landim  ${}^{a}$\footnote{e-mail: renan@fisica.ufc.br}  and R.N. Costa Filho  ${}^{a}$\footnote{e-mail: rai@fisica.ufc.br}}

\vspace{0,5cm}

 {\it ${}^a$Departamento de F\'{\i}sica, Universidade Federal do Cear\'{a}-
Caixa Postal 6030, Campus do Pici, 60455-760, Fortaleza, Cear\'{a}, Brazil. 
 }

\end{center}

\vspace{0.3cm}

\begin{abstract}

It is shown here that the zero mode of any form field  can be trapped to the brane using the model proposed by  Ghoroku and Nakamura. We start proven that the equations of motion can be obtained without splitting the field in even and odd parts. The massive and tachyonic cases are studied revealing that this mechanism only traps  the zero mode. The result is then generalized to thick branes. In this scenario, the use of a delta like interaction of the quadratic term is necessary leading to a ``mixed'' potential with singular and smooth contributions. It is also shown that all forms produces an effective theory in the brane without gauge fixing.  The existence of resonances with the transfer matrix method is then discussed. With this we analyze the resonances and look for peaks indicating the existence of unstable modes. Curiously no resonances are found in opposition of other models in the literature. Finally we find analytical solutions for arbitrary $p-$forms when a  specific kind of smooth 
scenario is 
considered.

\end{abstract}
\end{titlepage}

\newcommand{\e}{\mbox{e}}
\newcommand{\hip}[1]{\, _2F_1\left(#1\right)}

\section{Introduction}
After gravity's geometrization by Einstein in $1915$ several models raised to solve unification related problems  using gauge theory by the introduction of compact extra dimension, as proposed by Kaluza and Klein (KK) \cite{Bailin:1987jd}. In these models the extra dimensions are considered compact as a way to recover the four dimension world. Depending on the compactification and the number of dimensions different kinds of fields in the lower dimensional theory can be obtained \cite{Salam:1981xd}. For example, one can start with the gravity field in five dimensions $G_{MN}$, were capital latin indexes running from $1$ to $5$. In four dimensions, upon dimensional reduction we get  a gravity ($G_{\mu\nu}$), a vector ($G_{\mu 5}$) and a scalar field ($G_{55}$), with greek indexes running from $1$ to $4$. Similarly for the gauge the spectrum in four dimensions has a vector and a scalar field $A_M\to (A_{\mu};A_5)$. When considering $p-$form fields the dimensional reduction provides, $B_{M_{1}...M_{p}}\to (B_{\
\mu_{1}...\mu_{p}};B_{\mu_{1}...\mu_{p-1}5})$, 
i.e., a four dimensional $p$-form and a four dimensional $(p-1)$-form.  Therefore, the four dimensional vector field can be obtained from a five dimensional $1$-form or from a five dimensional $2$-form. In fact these models provide us a plethora of massive modes for all  above-mentioned cases, the zero mode being just a particular one. The need of a compact dimension is due to the fact that in this way the "escaping" of the fields to the extra dimensions becomes a small correction.

From a different point of view, a scenario considering our world as a shell has been proposed in \cite{Gogberashvili:1998iu} and further developed in \cite{Jardim:2011gg,Jardim:2013jy,Akama:2011wi}. Probably based in \cite{Gogberashvili:1998iu} Lisa Randall and Raman Sundrum (RS) proposed another scenario with four dimensional branes in a five dimensional anti-de Sitter (AdS) space. In this scenario two different models has been considered: In the RS type I  a compact space with two branes and a $Z_2$ symmetry is used, solving the hierarchy problem \cite{Randall:1999ee}. Being a model with compact dimension the dimensional reduction works in a very similar way to the KK model; In the RS type II model just one brane embedded in a large extra dimension space is considered. The extra dimension is curved by a warp factor such that the model has been considered as an alternative to compactification \cite{Randall:1999vf}.  As a model for large extra dimensions the issue of zero mode localization of 
fields is important. In fact in the last ten years the zero mode localization of gauge field has becomes a drawback in these model. This localization is necessary since in a four dimension space fields propagating into the bulk can not be observed. Moreover, it has been found that the zero mode of gravity and scalar fields are localized \cite{Randall:1999vf,Bajc:1999mh} in a positive tension brane. However, due to its conformal invariance the vector field is not localized, which is a serious problem for a realistic model. This problem has been studied in many ways. For example,  some authors introduced a dilaton coupling to solve the problem \cite{Kehagias:2000au}, and others like in \cite{Dvali:1996xe} proposed that a strongly coupled gauge theory in five dimensions can generate a massless photon in the brane. Moreover, studies using a topological mass term in the bulk has been introduced. But they were not able to generate a massless photon in the brane \cite{Oda:2001ux}. Most of these models introduces 
other fields or nonlinearities to the gauge field \cite{Chumbes:2011zt}.
As a solution for this, geometrical couplings has been proposed recently, as can be seen in Refs. \cite{Alencar:2014moa,Alencar:2014fga,Jardim:2014xla,Germani:2011cv}.

As a way to circumvent the lack of localization, the authors in \cite{Ghoroku:2001zu} introduced a mass term in five dimensions and a coupling with the brane. This gives a localized massless photon with an action given by
\begin{eqnarray}\label{delta}
S_{ A} &=& \int d^5X\sqrt{-G}\left(-{1\over 4}G^{MN}G^{PQ}Y_{MP}Y_{NQ}
               - {1\over 2}M^2 G^{MN}X_M X_N\right)   \cr
           && -c\int d^4x{1\over 2}\sqrt{-g}g^{MN}X_M X_N,
\end{eqnarray}
where $Y_{MN}=\partial_{[M} X_{N]}$. In this model the localization is obtained only for some values of the parameter $c$ and for a range in $M$. On the other hand, beyond the gauge field (one form) other forms can be considered. In $D-$dimensions one can in fact have the existence of any $p\leq D$. However, they can be considered in a unified way. The analysis of localizability of form fields has been considered in \cite{Kaloper:2000xa} where in $D-$dimensions only the forms with $p<(D-3)/2$ can be localized. This is a known result where in the absence of a topological obstruction the field strength of a $p-$form is dual to the $(D-p-2)-$form \cite{Duff:1980qv}. Using this the authors in \cite{Duff:2000se} found that in fact also for $p>(D-1)/2$ the field can be localized. It is important to point that in the models mentioned here the Hodge Duality is not valid since quadratic terms breaks this 
duality. The analysis of massive modes of $p-$forms has also been considered in the literature \cite{Alencar:2010mi,Alencar:2010hs,Alencar:2010vk,Landim:2010pq,Alencar:2013jm,Fu:2012sa}

After the work of RS several recent results have been developed based on the idea of thick membranes and its implications for brane-world physics \cite{Bazeia:2005hu,Bazeia:2004yw,Bazeia:2003qt,Liu:2009ve,Zhao:2009ja,Liang:2009zzf,Zhao:2010mk,Zhao:2011hg, Ahmed:2013lea, Guo:2011wr, Dzhunushaliev:2009va, Movahed:2007ps, Du:2013bx,Gremm:1999pj,Bazeia:2004dh,Bazeia:2006ef,Csaki:2000fc}. The advantage of these models is that the singularity generated by the brane is eliminated. In this scenario a transfer matrix method has been proposed to analyze resonances \cite{Landim:2011ki,Landim:2011ts,Alencar:2012en}.  However, when considering a smooth version of the model \cite{Ghoroku:2001zu}  we find that the $\delta(z)$ coupling in (\ref{delta})  is still necessary. With this we get the unexpected situation in which we have a ``mixing'' of a smooth and a singular potential. For some of the smooth cases we are able to find analytic solutions (for further analytic solutions see 
\cite{Cvetic:2008gu,Alencar:2012en,Landim:2013dja}).  When considering the resonances we must be careful with the mixing cited above. Because of this we need to modify our transfer matrix program to consider a delta like singularity in $z=0$. With this we analyze the resonances with the the transfer matrix method looking for peaks which indicate the existence of unstable modes \cite{Landim:2011ki}.

Here we show that the zero mode of any form field  can be trapped to the brane using the model proposed by  Ghoroku and Nakamura. We start proven that the equations of motion can be obtained without splitting the field in even and odd parts. The massive and tachyonic cases are studied revealing that this mechanism only traps  the zero mode. The result is then generalized to thick branes. This work is organized as follows: In Sec. II we use the Proca action with a coupling with the brane, similar to  \cite{Ghoroku:2001zu}, to localize a transversal massless photon in the brane. We use the definition of transversal and longitudinal parts of vector field to decouple the fields equation, instead impose parity in fields. We  find the coupling constant and the mass range that localizes the zero mode of reduced longitudinal vector field, leaving massive modes non-localized. We also study the zero and massive modes of reduced transversal vector field and scalar field. In Sec. III the same procedure used in previous 
section is used in a Kalb-Ramond field case. In this case the coupling constant and mass range is computed to localize the zero modes of reduced transversal Kalb-Ramond field and reduced gauge fields in the brane, while the massless mode of reduced longitudinal Kalb-Ramond field and its all massive modes are non-localized. In Sec. IV we generalize the previous results for a $p$-form field in a $D$-dimensional bulk. In this case we find the coupling constant and the range of mass parameter which localizes the zero mode of reduced longitudinal $p$-form. We find an link between $p$ and $D$ which localize the massless mode of $(p-1)$-form. We also studies the both massive cases and find a condition localizing the reduced transversal $p$-form. In Sec. V we use the procedure used in previous section for a smooth warp factor scenario. Unlike the case of thin membranes we find a fixation of coupling constant which localize all zero modes, for all $D$ and $p$. We study the massive modes numerically using the transfer 
matrix method to plot the transmission coefficient. In last section we discuss the conclusion and possibles consequences.

\section{Five Dimensional Proca Model in a Thin Brane Scenario}

As mentioned in the introduction, the action used in this paper is the following 
\begin{equation}\label{action}
 S_{5}=\int d^{5}X\sqrt{-g}\left(-\frac{1}{4}g^{MN}g^{PQ}Y_{MP}Y_{NQ}-\frac{1}{2}(M^{2}+c\delta(z))g^{MN}X_{M}X_{N}\right),
\end{equation}
this is the Proca action with a coupling term with the brane used in \cite{Ghoroku:2001zu}.
The parameter $c$, in (\ref{delta}), relates to above one by transformation $c \to \mu^{2} = (\sqrt{-g}/\sqrt{-g_{4}})c$; were $\mu$ is a four dimensional mass parameter and $g_{4}$ is the determinant
of induced four dimensional metric. The Randall-Sundrum metric in a conformal form, $g_{MN} = \e^{2A(z)}\eta_{MN}$ is used, where
$\eta_{MN} =\mbox{Diag.}(-1,1,1,1,1)$ and
\begin{equation}\label{RSmetric}
 A(z) = -\ln(k|z| +\beta),
\end{equation}
is the warp factor, which will be maintained in a generic way wherever possible to keep the generality. 
Taking the variation relative to $X_{N}$ we obtain the equation of motion
\begin{equation}\label{motion}
\partial_{M}(\sqrt{-g}g^{MO}g^{NP}Y_{OP}) -(M^{2}+c\delta(z))\sqrt{-g}g^{NP}X_{P} = 0.
\end{equation}
Although we have a massive gauge field, taking the divergence, the above equation implies
\begin{equation}\label{divergence}
 \partial((M^{2}+c\delta(z))\e^{3A}X_{5})+(M^{2}+c\delta(z))\e^{3A}\partial_{\mu}X^{\mu}=0,
\end{equation}
where $\partial$ means a derivative in $z$ and from now on all four dimensional indexes will be contracted with
$\eta^{\mu\nu}$. Fixing now $N=5$ in the equations of motion (\ref{motion}) we obtain
\begin{equation}\label{proca5}
\partial_{\mu}Y^{\mu5} -\e^{2A}(M^{2}+c\delta(z))X_{5}=0,
\end{equation}
 and for $N=\mu$
\begin{equation}\label{procanu}
 \e^{A}\partial_{\mu}Y^{\mu\nu}+\partial\left(\e^{A}Y^{5\nu}\right)-(M^{2}+c\delta(z))\e^{3A}X^{\nu}=0.
\end{equation}
The only way to associate this to a massive gauge field in four
dimension is  to have the condition $\partial_{\mu}X^{\mu}=0$ satisfied.
This is not true since we must have (\ref{divergence}). However,
we can split our field in two parts $X^{\mu}=X_{L}^{\mu}+X_{T}^{\mu}$,
where $L$ stands for longitudinal and $T$ stands for transversal
with
\begin{equation}
 X_{T}^{\mu}\equiv(\delta_{\nu}^{\mu}-\frac{\partial^{\mu}\partial_{\nu}}{\Box})X^{\nu}\;\;\;;\;\;\;X_{L}^{\mu}\equiv\frac{\partial^{\mu}\partial_{\nu}}{\Box}X^{\nu}.
\end{equation}
Next step is to show that the above equations can give effective
equations for a massive gauge field in four dimensions and a massive
scalar field (zero form) defined by $X_{5}=\phi$. The first thing
we observe is that (\ref{divergence}) will give a relation between
the scalar field and the longitudinal part of $X^{\mu}$. For the equations
of motion we use the following properties
\begin{equation}
 \partial_{\mu}Y^{\mu\nu}=\Box X_{T}^{\nu}\;\;\;;\;\;\;Y^{5\mu}=\partial X_{T}^{\mu}+\partial X_{L}^{\mu}-\partial^{\mu}\phi=\partial X_{T}^{\mu}+Y_{L}^{5\mu}
\end{equation}
and we get, from (\ref{proca5})
\begin{equation}\label{eqYLphi}
 \partial_{\mu}Y_{L}^{\mu5}-\e^{2A}(M^{2}+c\delta(z))\phi=0,
\end{equation}
and, from (\ref{procanu})
\begin{equation}\label{XTXL}
\e^{A}\Box X_{T}^{\nu}+\partial\left(\e^{A}\partial X_{T}^{\nu}\right)-(M^{2}+c\delta(z))\e^{3A}X_{T}^{\nu}+\partial(\e^{A}Y_{L}^{5\mu})-(M^{2}+c\delta(z))\e^{3A}X_{L}^{\nu}=0. 
\end{equation}
The longitudinal, the transversal and the scalar field ($X^{5}$) are coupled. 

For the scalar field the divergence equation (\ref{divergence}) can be used in (\ref{eqYLphi}) to give us, for $z \neq 0$
\begin{equation}
\Box\phi+\partial\left[\e^{-3A}\partial(\e^{3A}\phi)\right]-\e^{2A}M^{2}\phi=0 
\end{equation}
and the boundary condition
\begin{equation}
 c\phi(0,x) = \e^{-2A(0)}\left.\lim_{\epsilon \to 0}\e^{-3A}\partial(\e^{3A}\phi)\right|_{-\epsilon}^{\epsilon}.
\end{equation}
The two equations above can be rewritten as
\begin{equation}\label{phifull}
 \Box\phi+\partial[\e^{-3A}\partial(\e^{3A}\phi)]-\e^{2A}(M^{2}+c\delta(z))\phi=0,
\end{equation}
for all $z$.

At this point Ref. \cite{Ghoroku:2001zu} used an unnatural procedure to cancel the last two terms of (\ref{XTXL})and obtain
a mass equation for the transversal part of the vector field. This is done by imposing some parities to the fields
such that theses terms disappear. However, this is not necessary as can be seen: First note that the field $X_{T}^{\mu}$ satisfy
the traceless condition, and an equation for a massive gauge field should appear naturally no matter the
parity of the fields. The only way to have that is canceling the longitudinal
part by the scalar field part. In order to prove that, the definition of $X_{L}$ is used leading to
\begin{equation}
Y_{L}^{\mu5}=\frac{\partial^{\mu}}{\Box}\partial_{\nu}Y^{\nu5}= (M^{2}+c\delta(z))\e^{2A}\frac{\partial^{\mu}}{\Box}\phi,
\end{equation}
where in the last equation we have used equation (\ref{proca5}). Using now the divergence equation we get
\begin{equation}
\partial(\e^{A}Y_{L}^{5\mu})=-\frac{\partial^{\mu}}{\Box}\partial((M^{2}+c\delta(z))\e^{3A}\phi)=(M^{2}+c\delta(z))\e^{3A}\frac{\partial^{\mu}}{\Box}(\partial_{\nu}X^{\nu})=(M^{2}+c\delta(z))\e^{3A}X_{L}^{\mu}. 
\end{equation}
This term cancels the longitudinal part of (\ref{XTXL})leading to
\begin{equation}\label{XTfull}
\Box X_{T}^{\nu}+\e^{-A}\partial\left(\e^{A}\partial X_{T}^{\nu}\right)-(M^{2}+c\delta(z))\e^{2A}X_{T}^{\nu}=0. 
\end{equation}
Finally a set of decoupled equations is obtained governing the transversal part of gauge field, eq. (\ref{XTfull}), and the scalar field, eq. (\ref{phifull}). The vector field longitudinal part is linked to the scalar field (\ref{divergence}).

\begin{figure}[h!]
 \centering
 \includegraphics[scale=0.4]{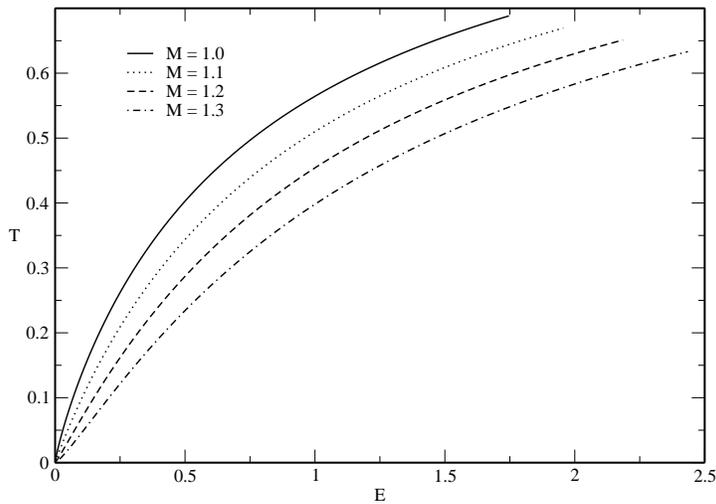}
 % Tdelta-vector.pdf: 792x612 pixel, 72dpi, 27.94x21.59 cm, bb=0 0 792 612
 \caption{Plot of the transmission coefficient as a function of energy, $E = m^{2}_{X}$, for different values of the Proca parameter $M$. We have fixed $ k = \beta = 1$.}
 \label{fig: tdeltavector}
\end{figure}

For the transversal part of $X^{\mu}$ a separation of variables $ X_{T}^{\nu}(z,x) = f(z)\tilde{X}_{T}^{\nu}(x)$ in (\ref{XTfull}) is used to obtain the following set of equations
\begin{eqnarray}
&& \Box\tilde{X}_{T}^{\nu}-m_{X}^{2}\tilde{X}_{T}^{\nu}=0,
\\&& \e^{-A}\left(\e^{A}f'(z) \right)'-(M^{2}+c\delta(z))\e^{2A}f(z) =-m_{X}^{2}f(z), \label{eqf}
\end{eqnarray}
where the prime means a derivative in $z$. The former is the equation of a reduced massive gauge field, while the later is the equation governing the localization factor $f(z)$. To put equation (\ref{eqf}) in a Schrödinger form the transformation $f(z) = \e^{-A/2}\psi(z)$ is used, with the effective potential
\begin{equation}
U(z)=\frac{1}{4}A'^{2}+\frac{1}{2}A''+(M^{2}+c\delta(z))\e^{2A},
\end{equation}
and using the Randall-Sundrum metric (\ref{RSmetric}) we can write
\begin{equation}
 U(z)=\frac{3k^{2}/4+M^{2}}{(k|z|+\beta)^{2}}-b_{1}\delta(z)
\end{equation}
where $b_{1}=k/\beta-c/\beta^{2}$. This potential provides the solution for the zero mode 
\begin{equation}\label{solf}
\psi = f_{0}(k|z|+\beta)^{1/2-\nu} + f_{1}(k|z|+\beta)^{1/2+\nu}
\end{equation}
where $f_{0}$ and $f_{1}$ are constants and $\nu=\sqrt{1+M^{2}/k^{2}}$. The boundary condition at $z=0$ impose the following relation
\begin{equation}
 \left(2k\beta(\nu -1) +c\right)f_{0} =  \left(2k\beta(\nu +1) -c\right)f_{1}.
\end{equation}
To get a desired localized solution we need fix $f_{1} = 0$, i.e., we need fix the free parameter $c$ as
\begin{equation}
c=-2k\beta(\nu-1),
\end{equation}
or, in terms of four dimensional mass parameter
\begin{equation}
 \mu^{2} = -2k(\nu-1).
\end{equation}
Therefore we see that any solution with $M^{2}>0$ will give a localized
solution. It is important to point that the four dimension mass parameter does not depends on $\beta$, is just fine-tuned with the five dimensional mass and the cosmological constant in bulk. For the massive modes we have a non-localized solution given by
\begin{equation}\label{psimas}
 \psi=(k|z|+\beta)^{1/2}[C_{1}J_{\nu}(m_{X}|z|+\beta m_{X}/k)+C_{2}Y_{\nu}(m_{X}|z|+\beta m_{X}/k)],
\end{equation}
where $C_{1}$ and $C_{2}$ are constants. To fit the boundary condition this constants must satisfy
\begin{equation}\label{condpm}
 C_{1} = C_{2}\frac{\beta m_{X} Y_{ \nu -1}(\beta m_{X}/k) - 2k\nu Y_{\nu}(\beta m_{X}/k) - 
  \beta m_{X}Y_{\nu+1}(\beta m_{X}/k)}{\beta m_{X} J_{\nu-1}(\beta m_{X}/k) + 2k\nu J_{\nu}(\beta m_{X}/k) -
   \beta m_{X} J_{\nu +1}(\beta m_{X}/k)}.
\end{equation}
The above condition do not allow $C_{1} = 0$ to obtain a localized solution. Then the massive modes are non-localized. To obtain more information
about massive modes one can evaluate the transmission coefficient. For this we will write the solution (\ref{psimas}) in the form
\begin{equation}
 \psi(z) = \left\lbrace\begin{matrix}E_{\nu}(-z)+\sigma F_{\nu}(-z) &,\;\mbox{for}\; z<0 \\ 
\gamma F_{\nu}(z)&,\;\mbox{for}\; z\geq0\end{matrix}\right.,
\end{equation}
where
\begin{eqnarray}
&&E_{\nu}(z) = \sqrt{\frac{\pi}{2}}(m_{X}z+\beta m_{X}/k)^{1/2}H^{(2)}_{\nu}(m_{X}z+\beta m_{X}/k)
\\&& F_{\nu}(z) = \sqrt{\frac{\pi}{2}}(m_{X}z+\beta m_{X}/k)^{1/2}H^{(1)}_{\nu}(m_{X}z+\beta m_{X}/k),
\end{eqnarray}
and $H^{(1)}_{\nu}$ and $H^{(2)}_{\nu}$ are the Hankel functions of first and second kind respectively. The boundary conditions at $z = 0$ leads to 
\begin{equation}
 \gamma =  \frac{W(E_{\nu},F_{\nu})(0)}{ 2F_{\nu}(0)F_{\nu}'(0) +  b_{1} F_{\nu}^{2}(0)},
\end{equation}
where $ W(E_{\nu},F_{\nu})(0) =  E_{\nu}(0)F_{\nu}'(0)-E_{\nu}'(0)F_{\nu}(0)$ is the Wronskian taking at $z=0$. This function is constant for the Schrödinger's equation and can be computed using the asymptotic behavior of the Hankel function. The transmission coefficient can be written in the form
\begin{equation}
T = |\gamma|^{2} = \frac{4m_{X}^{2}}{|2F_{\nu}(0)F_{\nu}'(0) +  b_{1} F_{\nu}^{2}(0)|^{2}}. 
\end{equation}
To illustrate the coefficient behavior Figure \ref{fig: tdeltavector} shows the transmission coefficient against the energy, $E = m_{X}^{2}$. The absence of peaks in the Figure indicates no unstable massive modes.
For tachyonic modes, taking $m_{X} \to im_{X}$ in (\ref{psimas}), we obtain a non-localized solution given by modified Bessel functions with a condition of coefficients similar  to (\ref{condpm}).

For the scalar field,using a separation of variables $\phi(z,x) = u(z)\tilde{\phi}(x)$, we obtain from (\ref{phifull}), the set of equations
\begin{eqnarray}
&& \Box\tilde{\phi}-m_{\phi}^{2}\tilde{\phi} =0,
\\&& [\e^{-3A}(\e^{3A}u(z))']'-\e^{2A}(M^{2}+c\delta(z))u(z)= -m_{\phi}^{2}u(z).\label{equ}
\end{eqnarray}
The first is the equation of reduced massive scalar field, while the second is the equation governing the localization factor $u(z)$. Using the transformation $u(z)=\e^{-3A(z)/2}\psi$
we can write eq. (\ref{equ}) in a Schrödinger form with a potential given by 
\begin{equation}
 U(z)=\frac{9}{4}A'^{2}-\frac{3}{2}A''+(M^{2}+c\delta(z))\e^{2A}
\end{equation}
or, for  Randall-Sundrum background
\begin{equation}
U(z)=\frac{3k^{2}/4+M^{2}}{(k|z|+\beta)^{2}}-\bar{b}_{1}\delta(z), 
\end{equation}
where $\bar{b}_{1}=-3k/\beta-c/\beta^{2}$. The above potential provides the same solution (\ref{solf}), with new constants $\bar{f}_{0}$ and $\bar{f}_{1}$, but now the boundary condition fix 
\begin{equation}
 \left(2k\beta(\nu -1) +c\right)\bar{f}_{0} =  \left(2k\beta(\nu +1) -c\right)\bar{f}_{1}.
\end{equation}
As we have fixed $c$ to localize the zero mode of vector field is not possible vanish the divergent part of scalar field solution. Therefore we can not to have both, the vector and the scalar field localized. The potential for scalar field is the same of vector field, changing only the boundary condition, the behavior of massive and tachyonic modes are the same, i.e, non-localized.

The longitudinal part of vector field can be found solving the divergence equation (\ref{divergence}). Separating the variables in the form $X_{L}^{\mu} = F(z)\tilde{X}_{L}^{\mu}$ we obtain

\begin{equation}
 F(z) = \mbox{sgn}(z)\left[F_{0}(1 -\nu)(k|z| +\beta)^{1-\nu} +F_{1}\left(1 +\nu\right)(k|z| +\beta)^{1+\nu}\right]
\end{equation}
where $F_{0}$ and $F_{1}$ are constants proportional to $\bar{f}_{0}$ and $\bar{f}_{1}$ respectively. In the same way that we can not vanish $\bar{f}_{1}$ due the boundary condition is not possible
vanish the divergent part of longitudinal vector field. This result show us that the longitudinal part of vector field is non-localized.

The five dimensional action (\ref{action}), which can be written using eqs. (\ref{divergence}), (\ref{eqf}) and (\ref{equ}) in the form
\begin{eqnarray}
S_{5} &=& \int\e^{A}f(z)^{2} dz\int d^{4}x\left[-\frac{1}{4}\tilde{Y}_{T\mu\nu}\tilde{Y}_{T}^{\mu\nu}  -\frac{1}{2}m_{X}^{2}\tilde{X}_{T\mu}\tilde{X}_{T}^{\mu}\right]+ \nonumber
\\&&+ \int \e^{A}u(z)^{2}dz\int d^{4}x\left[-\frac{1}{2} \partial_{\mu}\tilde{\phi}\partial^{\mu}\tilde{\phi} -\frac{1}{2} m_{\phi}^{2}\tilde{\phi}\tilde{\phi}\right]+ \nonumber
\\&&+\frac{1}{2}\int \e^{A} \left(\e^{-3A}(\e^{3A}u(z))'\right)'u(z)dz\int d^{4}x\tilde{\phi}\tilde{\phi}- \nonumber
\\&&- \int \e^{A}F^{2}dz\int d^{4}x\left[\frac{1}{4}\tilde{Y}_{L\mu\nu}\tilde{Y}_{L}^{\mu\nu} - \frac{1}{2}m_{L}^{2}\tilde{X}_{L\mu}\tilde{X}_{L}^{\mu}\right], \label{Ar1}
\end{eqnarray}
where $m_{L}$ is defined by
\begin{equation}
 \e^{-A}\left(\e^{A}F'(z) \right)'-(M^{2}+c\delta(z))\e^{2A}F(z) =-m_{L}^{2}F(z).
\end{equation}
The results obtained in this section shows that the above action reduces to standard action of massless gauge vector field 
\begin{equation}
 S_{4} =\int d^{4}x\left( -\frac{1}{2}\partial_{\mu}\tilde{X}_{T}^{\nu}\partial^{\mu}\tilde{X}_{T\nu}\right)
\end{equation}
on the brane if $\nu>1$. 
despite the $\tilde{X}_{T}^{\nu}$ field has zero divergence, due to gauge symmetry is restored on the brane, it is possible to define a field $\tilde{X}^{\nu}  = \tilde{X}_{T}^{\nu} +\partial^{\nu}\chi$ , so that
\begin{equation}
 S_{4} =\int d^{4}x\left( -\frac{1}{4}\tilde{Y}_{\mu \nu}\tilde{Y}^{\mu\nu}\right),
\end{equation}
where $\tilde{Y}_{\mu \nu}$ is the propagator of $\tilde{X}^{\nu}$. 
\section{The Kalb-Ramond Case}
Now we must generalize our result for the two form field. First of all we must remember that
upon dimensional reduction we are left with to kinds of terms, namely
a Kalb-Ramond in four dimensions $B_{\mu\nu}$ and a vector field
$B_{\mu5}$. In the case with gauge symmetry the vector contribution
could be canceled by a gauge choice, but here we are forced to
keep them. Therefore, we can visualize the interesting possibility
of having a Kalb-Ramond and a vector field localized in the membrane.
The action is  
\begin{equation}\label{S2form}
S_{5} =\int d^{5}x\sqrt{-g}\left[-\frac{1}{12}Y_{M_{1}M_{2}M_{3}}Y^{M_{1}M_{2}M_{3}}-\frac{1}{4}(M^{2}+c\delta(z))X_{M_{1}M_{2}}X^{M_{1}M_{2}}\right],
\end{equation}
where, like in 1-form case, the parameter $c$ relates to four dimensional mass by $ c = (\sqrt{-g_{4}}/\sqrt{-g})\mu^{2}$. The equations of motion are  
\begin{equation}\label{motion2form}
\partial_{M_{1}}\left[\sqrt{-g}g^{M_{1}M_{4}}g^{M_{2}M_{5}}g^{M_{3}M_{6}}Y_{M_{4}M_{5}M_{6}}\right]-(M^{2}+c\delta(z))\sqrt{-g}g^{M_{2}M_{5}}g^{M_{3}M_{6}}X_{M_{5}M_{6}}=0.
\end{equation}
Similarly to the one form case we get from the above equation the
condition
\begin{equation}\label{traceless2form}
\partial_{M_{1}}\left[(M^{2}+c\delta(z))\sqrt{g}g^{M_{1}M_{2}}g^{M_{3}M_{4}}X_{M_{2}M_{3}}\right]=0. 
\end{equation}
Now we can obtain the equations of motion by expanding eq. (\ref{motion2form}).
For $M_{2}=\mu_{2}$ and $M_{3}=\mu_{3}$ we obtain$ $
\begin{equation}\label{2formnu}
 \e^{-A}\partial_{\mu_{1}}Y^{\mu_{1}\mu_{2}\mu_{3}}+\partial(\e^{-A}Y^{5\mu_{2}\mu_{3}})-(M^{2}+c\delta(z))\e^{A}X^{\mu_{2}\mu_{3}}=0;
\end{equation}
and for $M_{3}=5$ we get
\begin{equation}\label{2form5}
 \partial_{\mu_{1}}Y^{\mu_{1}\mu_{2}5}-(M^{2}+c\delta(z))\e^{2A}X^{\mu_{2}}=0,
\end{equation}
where we have defined $X^{\mu} \equiv X^{\mu5}$ and, like in previous case, $\partial$ means a derivative in $z$ and from now on all four indexes will be contracted with $\eta^{\mu\nu}$. The divergence equation (\ref{traceless2form}), differently of the vector case, will
give rise to two equations. For $M_{4}=5$ we get $ \partial_{\mu}X^{\mu}=0$,
where we have used the previous definitions. Therefore we see that
the traceless condition for our vector field is naturally obtained upon
dimensional reduction. For $M_{4}=\mu_{4}$ we get 
\begin{equation}\label{traceless2formmu}
\partial((M^{2}+c\delta(z))\e^{A}X^{\mu_{4}})+\e^{A}(M^{2}+c\delta(z))\partial_{\mu_{1}}X^{\mu_{1}\mu_{4}}=0
\end{equation}
Just as in the case of the one form, here we have effective equations
that couples the Kalb Ramond and the Vector field. Before we proceed
to solve the equations we can further simplify them if we take the
longitudinal and transversal part of each field. As the vector field
already satisfy the traceless condition we just need to perform this
for the KR field by $X^{\mu_{1}\mu_{2}}=X_{L}^{\mu_{1}\mu_{2}}+X_{T}^{\mu_{1}\mu_{2}}$, defined
as
\begin{equation}
X_{T}^{\mu_{1}\mu_{2}}=X^{\mu_{1}\mu_{2}}+\frac{1}{\Box}\partial^{[\mu_{1}}\partial_{\nu_{1}}X^{\mu_{2}]\nu_{1}};\;\;\;\;X_{L}^{\mu_{1}\mu_{2}}=-\frac{1}{\Box}\partial^{[\mu_{1}}\partial_{\nu_{1}}X^{\mu_{2}]\nu_{1}}; 
\end{equation}
and observing that
\begin{equation}
 \partial_{\mu_{1}}Y^{\mu_{1}\mu_{2}\mu_{3}}=\square X_{T}^{\mu_{2}\mu_{3}};\;\;\;\; Y^{\mu_{1}\mu_{2}5}= Y^{\mu_{1}\mu_{2}5}_{L} +\partial X^{\mu_{1}\mu_{2}}_{T},
\end{equation}
the equations (\ref{2formnu}) and (\ref{2form5}) become
\begin{eqnarray}
&& e^{-A}\square X_{T}^{\mu_{2}\mu_{3}}+\partial(\e^{-A}\partial X_{T}^{\mu_{2}\mu_{3}})-(M^{2}+c\delta(z))\e^{A}X_{T}^{\mu_{2}\mu_{3}}+ \nonumber
 \\&&+\partial(\e^{-A}Y_{L}^{5\mu_{2}\mu_{3}})-(M^{2}+c\delta(z))\e^{A}X_{L}^{\mu_{2}\mu_{3}}=0 \label{eqXTXL2form}
\end{eqnarray}
and
\begin{equation}\label{eqYLphi2form}
\partial_{\mu_{1}}Y_{L}^{\mu_{1}\mu_{2}5}-(M^{2}+c\delta(z))\e^{2A}X^{\mu_{2}}=0,
\end{equation}
respectively.
Therefore, we see clearly from eq. (\ref{eqXTXL2form}) that we have a coupling between
the transversal part of the field, the longitudinal part and the gauge
field. Form eq. (\ref{eqYLphi2form}) we see that the gauge field is coupled to the
longitudinal part of the KR field. As in the case of the one form
field we should expect that we have to uncoupled effective massive
equation for the gauge fields $X_{T}^{\mu_{1}\mu_{2}}$ and $X^{\mu}$
since both satisfy the traceless condition in four dimensions. Lets
prove this now. First of all note that using $\partial_{\mu}X^{\mu}=0$
we can show that 
\begin{equation}
Y_{L}^{\mu_{1}\mu_{2}5}=-\frac{1}{\Box}\partial^{[\mu_{1}}\partial_{\nu}Y^{\mu_{2}]\nu} = (M^{2} +c\delta(z))\e^{2A}\frac{\partial^{[\mu_{1}}X^{\mu_{2}]}}{\Box},
\end{equation}
where in last equality we have used eq. (\ref{2form5}). Now we can use eq. (\ref{traceless2formmu}) in above identity to show that 
\begin{eqnarray}
\partial(\e^{-A}Y_{L}^{\mu_{1}\mu_{2}5})&=&\frac{\partial^{[\mu_{1}}}{\Box}\partial\left((M^{2} +c\delta(z))\e^{A}X^{\mu_{2}]}\right) =(M^{2} +c\delta(z))\e^{A}\frac{\partial^{[\mu_{1}}\partial_{\nu_{1}}X^{\mu_{2}]\nu_{1}}}{\Box} \nonumber
\\&=&(M^{2} +c\delta(z))\e^{A}X_{L}^{\mu_{1}\mu_{2}}
\end{eqnarray}
and this term cancels exactly the longitudinal part of the mass term.
Then we get the final form for the equation of motion
\begin{equation}\label{XT2full}
\e^{-A}\square X_{T}^{\mu_{1}\mu_{2}}+\partial(\e^{-A}\partial X_{T}^{\mu_{1}\mu_{2}})-(M^{2}+c\delta(z))\e^{A}X_{T}^{\mu_{1}\mu_{2}}=0 
\end{equation}
To decouple the vector field and the longitudinal part of KR field we can use eq.(\ref{traceless2formmu}) in (\ref{eqYLphi2form}) for $z \neq 0$
\begin{equation}
 \square X^{\mu_{2}}+\partial[\e^{-A}(\partial \e^{A}X^{\mu_{2}})] -M^{2}\e^{2A}X^{\mu_{2}} = 0
\end{equation}
and the boundary condition at $z=0$
\begin{equation}
cX^{\mu_{2}}(0,x) = \e^{-2A(0)}\lim_{\epsilon \to 0}\left.\e^{-A}\partial\left(\e^{A}X^{\mu_{2}}\right)\right|_{z=-\epsilon}^{z=\epsilon},
\end{equation}
or, summarizing for all $z$, 
\begin{equation}\label{phi2full}
\Box X^{\mu_{1}}+\partial\left[\e^{-A}\partial(\e^{A}X^{\mu_{1}})\right]-(M^{2} +\delta(z))\e^{2A}X^{\mu_{1}}=0.
\end{equation}
The set of decoupled equations (\ref{XT2full}) and (\ref{phi2full}) governs the transversal part of reduced Kalb-Ramond field and the reduced vector field respectively. Like in previous case
the longitudinal part of reduced Kalb-Ramond field  is linked to vector field by relation (\ref{traceless2formmu}). 

Beginning with transversal part of 2-form field, we impose the separation of variables in the form $ X_{T}^{\mu_{1}\mu_{2}}(z,x) = f(z)\tilde{X}_{T}^{\mu_{1}\mu_{2}}(x)$ in (\ref{XT2full}) to obtain the following set of equations
\begin{eqnarray}
&&  \square \tilde{X}_{T}^{\mu_{1}\mu_{2}}-m_{X}^{2}\tilde{X}_{T}^{\mu_{1}\mu_{2}}=0,
\\&& (\e^{-A}f'(z))'-(M^{2}+c\delta(z))\e^{A}f(z) = -m_{X}^{2}\e^{-A}f(z), \label{eqf2}
\end{eqnarray}
where the prime means a derivative in $z$. The first equation shows that $\tilde{X}_{T}^{\mu_{1}\mu_{2}}$ is a massive four dimensional form, while the second governs the localization factor $f(z)$.
To transform eq. (\ref{eqf2}) in a Schrödinger's equation we must  make $f(z)=\e^{A/2}\psi(z)$. Leading to the potential
\begin{equation}\label{pot2}
U(z)=\frac{A'^{2}}{4}-\frac{A''}{2}+(M^{2}+c\delta(z))\e^{2A} = k^{2}\frac{-\frac{1}{4}+M^{2}/k^{2}}{(k|z|+\beta)^{2}}-b_{2}\delta(z),
\end{equation}
with  $b_{2}=-k/\beta -c/\beta^{2}$. The zero mode solution is given by
\begin{equation}\label{solf2}
\psi= f_{0}(k|z|+\beta)^{1/2-\nu} +f_{1}(k|z|+\beta)^{1/2+\nu},
\end{equation}
where $\nu = M/k$, $f_{0}$ and $f_{1}$ are constants. Due to the boundary condition this constants must satisfy
\begin{equation}
\left(2k\beta\nu +c\right)f_{0} =\left(2k\beta\nu-c\right)f_{1}.
\end{equation}
To vanish the divergent part we can fix the free parameter as $c=-2k\nu\beta$, or in terms of four dimensional mass parameter
\begin{equation}
 \mu^{2} = -2k\nu.
\end{equation}
To be localized we must impose that $\nu >1$, i.e. $M/k > 1$. Like in $1$-form case, the fixation of four dimensional mass parameter does not depends on $\beta$.

\begin{figure}[h!]
 \centering
 \includegraphics[scale=0.4]{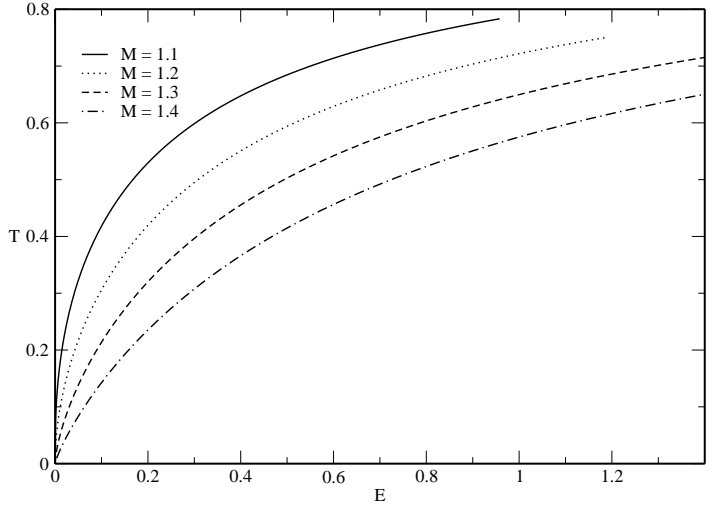}
 % Tdelta-KR.pdf: 792x612 pixel, 72dpi, 27.94x21.59 cm, bb=0 0 792 612
 \caption{Plot of transmission coefficient as function of energy, $E = m^{2}_{X}$, for different values of mass parameter $M$. We have fixed $ k = \beta = 1$.}
 \label{fig: tdeltaKR}
\end{figure}

For the massive case Eq (\ref{eqf2}) provides the solution 
\begin{equation}\label{psimas2form}
 \psi(z) =(k|z|+\beta)^{1/2}[C_{1}J_{\nu}(m_{X}|z|+\beta m_{X}/k)+C_{2}Y_{\nu}(m_{X}|z|+\beta m_{X}/k)],
\end{equation}
where $C_{1}$ and $C_{2}$ are constants. This solution is the same as the vector case, eq. (\ref{psimas}). Now the boundary conditions impose that
\begin{equation}
C_{1} = C_{2}\frac{\beta m_{X} Y_{\nu-1}(\beta m_{X}/k) -2\nu k Y_{\nu}(\beta m_{X}/k) - 
     \beta m_{X} Y_{\nu+1}(\beta m_{X}/k)}{\beta m_{X} J_{\nu-1}(\beta m_{X}/k) +2\nu k J_{\nu}(\beta m_{X}/k) -
   \beta m_{X}  J_{ \nu +1}(\beta m_{x}/k)}
\end{equation}
Like in vector case, the above condition does not allow us to find a localized solution for massive modes. Since the solution is the same of the previous case we can make the same procedure to obtain the transmission coefficient
\begin{equation}
T  = \frac{4m_{X}^{2}}{|2F_{\nu}(0)F_{\nu}'(0) +  b_{2} F_{\nu}^{2}(0)|^{2}}, 
\end{equation}
which is illustrated in figure \ref{fig: tdeltaKR}. The behavior of $T$ do not show peaks, indicating, again, no unstable massive modes.
For tachyonic modes, making $m_{X} \to im_{X}$ in (\ref{psimas2form}), we obtain a nonlocalized solution given by modified Bessel functions with a correspondent condition for coefficients. 

Now we can analyze the localizability of the vector field. Performing the separation of variables $X^{\mu_{1}}=u(z)\tilde{X}^{\mu_{1}}(x)$
we get, from (\ref{phi2full}), the set of equations
\begin{eqnarray}
&& \Box \tilde{X}^{\mu_{2}}-m_{1}^{2}\tilde{X}^{\mu_{2}}=0,
\\&& \left(\e^{-A}(\e^{A}u(z))'\right)'-\left(M^{2} +c\delta(z)\right)\e^{2A}u(z)= -m_{1}^{2}u(z).\label{equ2}
\end{eqnarray}
Like in previous cases the first equation shows that $\tilde{X}^{\mu_{2}}$ is a massive four-dimensional gauge field and the second equation provides the localization factor $u(z)$.  To transform the eq. (\ref{equ2}) in a Schrödinger form we need to make $u(z) = \e^{-A/2}\psi$. The potential obtained after this transformation is
\begin{equation}
U(z) = \frac{A'^{2}}{4}-\frac{A''}{2}+(M^{2}+c\delta(z))\e^{2A} = k^{2}\frac{-\frac{1}{4}+M^{2}/k^{2}}{(k|z|+\beta)^{2}}-{b}_{2}\delta(z)
\end{equation}
The potential is the same as (\ref{pot2}),
and give us the same solution to the zero mode, with new constants $\bar{f}_{0}$ and $\bar{f}_{1}$. Due the metric factor  we must have
the condition $\nu>2$, i.e., $M/k >2$ and therefore we can have both fields localized.
Whereby the potential for $1$-form is the same of $2$-form the behavior of massive and tachyonic modes are the same, i.e, non-localized.

The longitudinal part of $2$-form can be found solving eq. (\ref{traceless2formmu}). Imposing the separation of variables in the form  $X^{\mu_{1}\mu_{2}}_{L}(z,x) = F(z)\tilde{X}^{\mu_{1}\mu_{2}}_{L}(x)$ we
obtain
\begin{equation}
 F(z) =  F_{0}\mbox{sgn}(z)\left(2 -2\nu\right)(k|z| +\beta)^{-\nu} 
\end{equation}
where $F_{0}$ is a constant proportional to $\bar{f}_{0}$. Since the divergent part of vector field vanishes, the longitudinal part of Kalb-Ramond in localized if $\nu>1$.
The five-dimensional action (\ref{S2form}) can be write, using (\ref{eqYLphi2form}), (\ref{eqf2}) and (\ref{equ2}), in the form 
\begin{eqnarray}
S_{5} &=& \int\e^{-A}f(z)^{2} dz\int d^{4}x\left[-\frac{1}{12}\tilde{Y}_{T\mu_{1}\mu_{2}\mu_{3}}\tilde{Y}_{T}^{\mu_{1}\mu_{2}\mu_{2}}  -\frac{1}{4}m_{X}^{2}\tilde{X}_{T\mu_{2}\mu_{3}}\tilde{X}_{T}^{\mu_{2}\mu_{3}}\right]+ \nonumber
\\&&+ \int \e^{-A}u(z)^{2}dz\int d^{4}x\left[-\frac{1}{4} \tilde{Y}^{\mu_{1}\mu_{2}}\tilde{Y}_{\mu_{1}\mu_{2}} -\frac{1}{4} m_{1}^{2}\tilde{X}_{\mu_{2}}\tilde{X}^{\mu_{2}}\right]+ \nonumber
\\&& + \int \e^{-A}F^{2}dz\int d^{4}x\left[-\frac{1}{12}\tilde{Y}_{L\mu_{1}\mu_{2}\mu_{3}}\tilde{Y}_{L}^{\mu_{1}\mu_{2}\mu_{3}} - \frac{1}{4}m_{L}^{2}x\tilde{X}_{L\mu_{1}\mu_{2}}\tilde{X}_{L}^{\mu_{1}\mu_{2}}\right] +\nonumber
\\&&+\frac{1}{4}\int \e^{-A} \left(\e^{-A}(\e^{A}u(z))'\right)'u(z)dz\int d^{4}x\tilde{X}_{\mu_{2}}\tilde{X}^{\mu_{2}}, \label{Ar2}
\end{eqnarray}
where
\begin{equation}
\e^{A}\left(\e^{-A}F'\right)' -(M^{2}+c\delta(z))\e^{2A}F(z) = -m_{L}^{2}F(z).
\end{equation}
The results obtained in this section show that above action reduces to the action of massless Kalb-Ramond field plus a longitudinal massless 2-form field
\begin{eqnarray}
S_{4} &=& \int d^{4}x\left[-\frac{1}{12}\tilde{Y}_{\mu_{1}\mu_{2}\mu_{3}}\tilde{Y}^{\mu_{1}\mu_{2}\mu_{2}} \right],
\end{eqnarray}
on the brane if $1<\nu \leq 2$; and the above action plus a  vector field 
\begin{eqnarray}
S_{4} &=& \int d^{4}x\left[-\frac{1}{12}\tilde{Y}_{\mu_{1}\mu_{2}\mu_{3}}\tilde{Y}^{\mu_{1}\mu_{2}\mu_{2}} -\frac{1}{4} \tilde{Y}^{\mu_{1}\mu_{2}}\tilde{Y}_{\mu_{1}\mu_{2}} \right],
\end{eqnarray}
on the brane if $\nu >2$, where $\tilde{Y}_{\mu_{1}\mu_{2}\mu_{3}}$ is the propagator of the field $ \tilde{X}_{\mu_{1}\mu_{2}} \equiv \tilde{X}_{T\mu_{1}\mu_{2}} +\tilde{X}_{L\mu_{1}\mu_{2}}$.
The effective Kalb-Ramond field in four dimensions is composed by the transverse and longitudinal parts of the field from the field in five dimensions. Thus there is no gauge fixing.

\section{The $p-$form Case}
Now we try to localize any $p-$form field in a $(D-1)$-brane. The action is given by
\begin{eqnarray}
S_{D}&=&\int d^{D}x\sqrt{-g}\left[-\frac{1}{2(p+1)!}Y_{M_{1}...M_{p+1}}Y^{M_{1}...M_{p+1}}-\right. \nonumber
\\&&-\left.\frac{1}{2p!}(M^{2}+c\delta(z))X_{M_{2}...M_{p+1}}X^{M_{2}...M_{p+1}}\right],
\end{eqnarray}
where the parameter $c$ relates to the $(D-1)$-mass by $ c = (\sqrt{-g_{D-1}}/\sqrt{-g})\mu^{2}$, where $g_{D-1}$ is the determinant of induced $(D-1)$-metric.
The equations of motion are given by
\begin{equation}\label{motionpform}
\partial_{M_{1}}[\sqrt{-g}g^{M_{1}N_{1}}...g^{M_{p+1}N_{p+1}}Y_{N_{1}...N_{p+1}}]-(M^{2}+c\delta(z))\sqrt{-g}g^{M_{2}N_{2}}...g^{M_{p+1}N_{p+1}}X_{N_{2}...N_{p+1}}=0. 
 \end{equation}
Similarly to the one form case we get from the above equation the condition
\begin{equation}\label{divpform}
 (M^{2}+c\delta(z))\e^{(D-2p)A}\partial^{\nu_{2}}X_{\nu_{2} N_{3}...N_{p+1}} + \partial\left[(M^{2}+c\delta(z))\e^{(D-2p)A}X_{5N_{3}...N_{p+1}}\right]=0.
\end{equation}
where, like in previous sections, $\partial$ means a derivative with $z$ and from now on all $(D-1)$-dimensional indexes will be contracted with
$\eta^{\mu\nu}$. Now we can obtain the equations of motion by expanding eq (\ref{motionpform}).
We are going to have just two kinds of terms, when none of the
indexes is $5$, giving
\begin{eqnarray}
&&\e^{(D-2(p+1))A}\partial_{\mu_{1}}[Y^{\mu_{1}\mu_{2}...\mu_{p+1}}]+\partial(\e^{(D-2(p+1))A}Y^{5\mu_{2}...\mu_{p+1}})-\nonumber
\\&&-(M^{2}+c\delta(z))\e^{(D-2p)A}X^{\mu_{2}...\mu_{p+1}}=0; \label{pformnu}
\end{eqnarray}
and when one of the indexes is $5$ we get
\begin{equation}\label{pform5}
\partial_{\mu_{1}}Y^{\mu_{1}\mu_{2}...\mu_{p}5} -(M^{2}+c\delta(z))\e^{2A}X^{\mu_{2}...\mu_{p}5}=0. 
\end{equation}
The divergence equation (\ref{divpform}), differently of the vector case, will
give rise to two equations. For one index with $5$ we get $ \partial_{\mu_{1}}X^{\mu_{1}...\mu_{p-1}}=0$,
where we have used our previous definitions and $ X^{\mu_{1}...\mu_{p-1}5} \equiv X^{\mu_{1}...\mu_{p-1}}$. Therefore we see that
the traceless condition for our vector field is naturally obtained upon
dimensional reduction. For none indexes with $5$ we get 
\begin{equation}\label{tracelesspform}
\partial((M^{2}+c\delta(z))\e^{(D-2p)A}X^{\mu_{1}...\mu_{p-1}})+(M^{2}+c\delta(z))\e^{(D-2p)A}\partial_{\mu_{p}}X^{\mu_{1}...\mu_{p}}=0 
\end{equation}
First of all, due the gauge symmetry has been broken in $D$-dimensional action, we must divide the $p$-form in transversal and longitudinal parts. As before, we will define these parts as
\begin{equation}
X_{T}^{\mu_{1}...\mu_{p}}\equiv X^{\mu_{1}...\mu_{p}}+\frac{(-1)^{p}}{\Box}\partial^{[\mu_{1}}\partial_{\nu_{1}}X^{\mu_{2}...\mu_{p}]\nu_{1}};\;\;\;\;X_{L}^{\mu_{1}...\mu_{p}}\equiv \frac{(-1)^{p-1}}{\Box}\partial^{[\mu_{1}}\partial_{\nu_{1}}X^{\mu_{2}...\mu_{p}]\nu_{1}}. 
\end{equation}
Observing that
\begin{equation}
\partial_{\mu_{1}}Y^{\mu_{1}\mu_{2}...\mu_{p}}=\square X_{T}^{\mu_{2}...\mu_{p}};\;\;\;\;Y^{5\mu_{1}...\mu_{p}}=Y_{L}^{5\mu_{1}...\mu_{p}}+\partial X_{T}^{\mu_{1}...\mu_{p}} 
\end{equation}
we can write the equation (\ref{pformnu}) as
\begin{eqnarray}
&& \e^{(D-2(p+1))A}\square X_{T}^{\mu_{1}...\mu_{p}}+\partial(\e^{(D-2(p+1))A}\partial X_{T}^{\mu_{1}...\mu_{p}})-(M^{2}+c\delta(z))\e^{(D-2p)A}X_{T}^{\mu_{2}...\mu_{p+1}}+ \nonumber
\\&&+\partial(\e^{(D-2(p+1))A}Y_{L}^{5\mu_{1}...\mu_{p}})-(M^{2}+c\delta(z))\e^{(D-2p)A}X_{L}^{\mu_{1}...\mu_{p}}=0 \label{eqXTXLpform}
\end{eqnarray}
and (\ref{pform5}) as
\begin{equation}\label{eqYLphipform}
\Box X^{\mu_{2}...\mu_{p}}-(M^{2}+c\delta(z))\e^{2A}X^{\mu_{2}...\mu_{p}}=0. 
\end{equation}

Therefore we see clearly from eq. (\ref{eqXTXLpform}) that we have a coupling between
the transversal part of the field, the longitudinal part and the gauge
field. Form eq. (\ref{eqYLphipform}) we see that the gauge field is coupled to the
longitudinal part of the KR field. As in the case of the one form
field we should expect that we have to uncoupled effective massive
equation for the gauge fields $X_{T}^{\mu_{1}\mu_{2}}$ and $X^{\mu}$
since both satisfy the trace less condition in four dimensions. Lets
prove this now. First of all note that using $\partial_{\mu}X^{\mu}=0$
we can show that 

\begin{equation}
 Y^{\mu_{1}...\mu_{p}}=\frac{(-1)^{p-1}}{\Box}\partial^{[\mu_{1}}\partial_{\nu}Y^{\mu_{2}...\mu_{p}]\nu},
\end{equation}
and we get an identity similar to that for the gauge field
\begin{eqnarray}
Y_{L}^{\mu_{1}...\mu_{p}5}&&=(-1)^{p}\partial X_{L}^{\mu_{1}...\mu_{p}}+(-1)^{p}Y^{\mu_{1}...\mu_{p}}\nonumber
\\&&=\frac{(-1)^{p}}{\Box}\left[(-1)^{p-1}\partial\partial^{[\mu_{1}}\partial_{\nu}X^{\mu_{2}...\mu_{p}]\nu}+\partial^{[\mu_{1}}\partial_{\nu}Y^{\mu_{2}...\mu_{p}]\nu}\right] \nonumber
\\&&=\frac{(-1)^{p-1}}{\Box}\partial^{[\mu_{1}}\partial_{\nu}Y^{\mu_{2}...\mu_{p}]\nu5} = \frac{(M^{2}+c\delta(z))\e^{2A}}{\Box}\partial^{[\mu_{1}}X^{\mu_{2}...\mu_{p}]},
\end{eqnarray}
where in the last equation we have used equation (\ref{pform5}). Using now the divergence equation (\ref{tracelesspform}) we obtain
\begin{eqnarray}
\partial\left(\e^{(D-2(p+1))A}Y_{L}^{\mu_{1}...\mu_{p}5}\right) &&= \frac{\partial^{[\mu_{1}}}{\Box}\partial\left(\e^{(D-2p)A}(M^{2}+c\delta(z))X^{\mu_{2}...\mu_{p}]}\right) \nonumber
\\&&=-\e^{(D-2p)A}(M^{2}+c\delta(z))\frac{\partial^{[\mu_{1}}}{\Box}\partial_{\mu}X^{\mu_{2}...\mu_{p}]\mu}\nonumber
\\&&=(-1)^{p}(M^{2}+c\delta(z))\e^{(D-2p)A}X_{L}^{\mu_{1}...\mu_{p}} .
\end{eqnarray}
This term exactly cancels the longitudinal part of (\ref{eqXTXLpform}), and we finally get the decoupled equation of motion for transversal part of $p$-form
\begin{equation}\label{XTpfull}
 \e^{(D-2(p+1))A}\square X_{T}^{\mu_{1}...\mu_{p}}+\partial(\e^{(D-2(p+1))A}\partial X_{T}^{\mu_{1}...\mu_{p}})-(M^{2}+c\delta(z))\e^{(D-2p)A}X_{T}^{\mu_{1}...\mu_{p}}=0.
\end{equation}
To decoupled the $(p-1)$-form and the longitudinal part of $p$-form we can use eq. (\ref{tracelesspform}) in (\ref{pform5}) for $z \neq 0$ 
\begin{equation}
 \Box X^{\mu_{2}...\mu_{p}}+\partial\left(\e^{-(D-2p)A}\partial\left(\e^{(D-2p)A}X^{\mu_{2}...\mu_{p}}\right)\right)-M^{2}\e^{2A}X^{\mu_{2}...\mu_{p}}=0
\end{equation}
with the boundary condition 
\begin{equation}
 cX^{\mu_{2}...\mu_{p}}(0,x) =\e^{-2A(0)} \lim_{\epsilon \to 0}\left.\e^{-(D-2p)A}\partial\left(\e^{(D-2p)A}X^{\mu_{2}...\mu_{p}}\right)\right|_{z=-\epsilon}^{z=\epsilon},
\end{equation}
or, summarizing for all $z$, 
\begin{equation}\label{phipfull}
\Box X^{\mu_{1}...\mu_{p-1}}+\partial(\e^{-(D-2p)A}\partial(\e^{(D-2p)A}X^{\mu_{1}...\mu_{p-1}}))-(M^{2} +\delta(z))\e^{2A}X^{\mu_{1}...\mu_{p-1}}=0.
\end{equation}
Finally, we found the set of decoupled equations which governs the transversal part of $p$-form and $(p-1)$-form; eqs. (\ref{XTpfull}) and (\ref{phipfull}), respectively. The longitudinal
part of $p$-form keep coupled with lowest order form by (\ref{eqYLphipform}).

To solve eq. (\ref{XTpfull}) we impose the separation of variables in the form $ X_{T}^{\mu_{1}...\mu_{p}}(z,x) = f(z)\tilde{X}_{T}^{\mu_{1}...\mu_{p}}(x)$ to obtain
\begin{eqnarray}
&&  \square \tilde{X}_{T}^{\mu_{1}...\mu_{p}}-m_{X}^{2}\tilde{X}_{T}^{\mu_{1}...\mu_{p}}=0,
\\&& (\e^{(D-2(p+1))A}f'(z))'-(M^{2}+c\delta(z))\e^{(D-2p)A}f(z) =-m_{X}^{2}\e^{(D-2(p+1))A}f(z), \label{eqfp}
\end{eqnarray}
where primes means a derivative with respect to $z$.
Now, making $f(z) = \e^{-(D-2(p+1))A/2}\psi$, we can write eq. (\ref{eqfp}) in a Schrödinger form with potentials given by
\begin{equation}\label{potp}
U(z)= \frac{\alpha_{p}^{2}}{4}A'^{2}+\frac{\alpha_{p}}{2}A''+(M^{2}+c\delta(z))\e^{2A} = k^{2}\frac{\frac{\alpha_{p}^{2}}{4}+\frac{\alpha_{p}}{2}+M^{2}/k^{2}}{(k|z|+\beta)^{2}}-b_{p}\delta(z)
\end{equation}
with $\alpha_{p}=D-2(p+1)$ and $b_{p}=k\alpha_{p}/\beta -c/\beta^{2}$. The zero mode solution is given by
\begin{equation}\label{solfp}
\psi= f_{0}(k|z|+\beta)^{1/2-\nu} +f_{1}(k|z|+\beta)^{1/2+\nu},
\end{equation}
where
\begin{equation}\label{nup}
\nu = \sqrt{ \frac{1}{4} +\frac{\alpha_{p}^{2}}{4}+\frac{\alpha_{p}}{2}+M^{2}/k^{2}},
\end{equation}
and $f_{0}$ and $f_{1}$ are constants. Due the boundary condition this constants must satisfy
\begin{equation}
\left(k\beta(2\nu -\alpha_{p} -1) +c\right)f_{0} =\left(k\beta(2\nu +\alpha_{p}+1)-c\right)f_{1} .
\end{equation}
To vanish the divergent part we can fix the free parameter as $c=-k(2\nu-1-\alpha_{p})\beta$, or in terms of $(D-1)$-dimensional mass parameter
\begin{equation}
\mu^{2} = -k(2\nu-1-\alpha_{p}).
\end{equation}
This result generalize the fact that the parameter $\mu$ does not depends on $\beta$, is only fine-tuned with the five dimensional mass and the cosmological constant in bulk
. To be localized we must impose that $\nu >1$, i.e.
\begin{equation}\label{mmin}
M^{2}/k^{2} > -\frac{(\alpha_{p} +3)(\alpha_{p} -1)}{4}
\end{equation}
The above result show us that a localized solution can be found without massive term if $D>2p+3$ or $D< 2p-1$.

For the massive case the eq. (\ref{eqfp}) provides the solution
\begin{equation}\label{psimaspform}
 \psi(z) =(k|z|+\beta)^{1/2}[C_{1}J_{\nu}(m_{X}|z|+\beta m_{X}/k)+C_{2}Y_{\nu}(m_{X}|z|+\beta m_{X}/k)],
\end{equation}
where $C_{1}$ and $C_{2}$ are constants . This solution is the same of 1-form and 2-form case, eqs. (\ref{psimas}) and (\ref{psimas2form}). Now the boundary conditions impose that
\begin{equation}
C_{1} = C_{2}\frac{\beta m_{X} Y_{\nu-1}(\beta m_{X}/k) -2\nu k Y_{\nu}(\beta m_{X}/k) - 
     \beta m_{X} Y_{\nu+1}(\beta m_{X}/k)}{\beta m_{X} J_{\nu-1}(\beta m_{X}/k) +2\nu k J_{\nu}(\beta m_{X}/k) -
   \beta m_{X}  J_{ \nu +1}(\beta m_{x}/k)}
\end{equation}
Like in previous cases, the above condition do not allow us to find a localized solution for massive modes. The above result show us that the solution of massive modes are the same, independent of the degree of the form. 
So that the transmission coefficient will differ only due to the boundary condition, i.e.,
\begin{equation}
T  = \frac{4m_{X}^{2}}{|2F_{\nu}(0)F_{\nu}'(0) +  b_{p} F_{\nu}^{2}(0)|^{2}}.
\end{equation}
\begin{figure}[h!]
 \centering
 \includegraphics[scale=0.4]{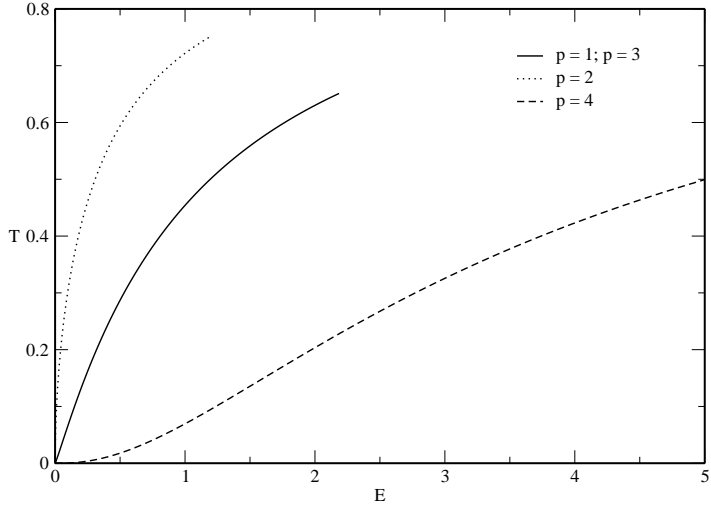}
 % Tdelta-p.pdf: 792x612 pixel, 72dpi, 27.94x21.59 cm, bb=0 0 792 612
 \caption{Plot of transmission coefficient as function of energy, $E = m^{2}_{X}$, for different $p$-forms in five dimensions. We have fixed $ k = \beta = 1$ and $M = 1.2$.}
 \label{fig: tdeltap}
\end{figure}
The transmission coefficient was plotted in fig. \ref{fig: tdeltap} as function of energy for some $p$-forms and do not show peaks, indicating no unstable massive modes.  

For tachyonic modes, making $m_{X} \to im_{X}$ in (\ref{psimaspform}), we obtain a non-localized solution given by modified Bessel functions with a correspondent condition for coefficients. 

For the $(p-1)$-form we have, imposing the separation of variables $X_{\mu_{2}...\mu_{p}}(z,x) = u(z)\tilde{X}_{\mu_{2}...\mu_{p}}(x)$, from (\ref{phipfull}) the set of equations
\begin{eqnarray}
&& \Box \tilde{X}^{\mu_{2}...\mu_{p}}-m_{p-1}^{2}\tilde{X}^{\mu_{2}...\mu_{p}}=0, 
\\&& \left(\e^{-(D-2p)A}(\e^{(D-2p)A}u(z))'\right)'-\left(M^{2} +c\delta(z)\right)\e^{2A}u(z)=-  m_{p-1}^{2}u(z).\label{equp}
\end{eqnarray}
To transform the eq. (\ref{equp}) in a Schrödinger form we need to make $u(z) = \e^{-(D-2p)A/2}\psi$. The potential obtained after this transformation is
\begin{equation}\label{potp-1}
U(z)=\frac{\alpha_{p-1}^{2}}{4}A'^{2}-\frac{\alpha_{p-1}}{2}A''+(M^{2}+c\delta(z))\e^{2A}= k^{2}\frac{\frac{\alpha_{p}^{2}}{4}+\frac{\alpha_{p}}{2}+M^{2}/k^{2}}{(k|z|+\beta)^{2}}+ \left(b_{p-1} +\frac{2c}{\beta^{2}}\right)\delta(z). 
\end{equation}
As a finite part of potential is the same as (\ref{potp}) the solution is the same as (\ref{solfp}) with new constants $\bar{f}_{0}$ and $\bar{f}_{1}$. Now the boundary condition at $z=0$ imposes the link between the constants
\begin{equation}
 \left(k\beta(1 -2\nu -\alpha_{p-1})-c\right)\bar{f}_{0} = \left(k\beta(1 +2\nu -\alpha_{p-1})-c\right)\bar{f}_{1},
\end{equation}
or, replacing $c$, found to vanish the divergent part of (\ref{solfp}), and $\alpha_{p-1}$
\begin{equation}
 \left[(2p+1)- D\right]\bar{f}_{0} = \left[ 2\nu -D+(2p+1)\right]\bar{f}_{1}.
\end{equation}
This result show us that a convergent solution can be obtained for $ D = 2p +1$, with an additional condition $\nu>2$, i.e., $M^{2}/k^{2} >-(\alpha_{p}-3)(\alpha_{p} +5)/4$ . For example, the cases treated before, with $D=5;p=2$ satisfy this condition. The potential for $(p-1)$-form is the same of $p$-form, changing only the boundary condition, the behavior of massive and tachyonic modes are the same, i.e, non-localized.

The longitudinal part of $p$-form can be found solving eq. (\ref{tracelesspform}). Imposing the separation of variables in the form  $X^{\mu_{1}...\mu_{p}}_{L}(z,x) = F(z)\tilde{X}^{\mu_{1}...\mu_{p}}_{L}(x)$ we
obtain
\begin{eqnarray}
 F(z) &=& \mbox{sgn}(z)\left[F_{0}\left(\alpha_{p} +3 -2\nu\right)(k|z| +\beta)^{(\alpha_{p}+1-2\nu)/2} \right.+ \nonumber
\\&&\left. +F_{1}\left(\alpha_{p} +3 +2\nu\right)(k|z| +\beta)^{(\alpha_{p}+1+2\nu)/2}\right]
\end{eqnarray}
where $F_{0}$ and $F_{1}$ are constants proportional to $\bar{f}_{0}$ and $\bar{f}_{1}$ respectively. Like in previous cases is possible to take $F_{1}$ out only if $\bar{f}_{1}$ vanishes, in this
case the longitudinal $p$-form will be localized if $\nu>1$.
The the $D$-dimensional action can be written, using (\ref{pformnu}), (\ref{eqfp}) and (\ref{equp}), in the form
\begin{small}
\begin{eqnarray}
S_{D} &=& \int\e^{\alpha_{p}A}f(z)^{2} dz\int d^{D-1}x\left[-\frac{1}{2(p+1)!}\tilde{Y}_{T\mu_{1}...\mu_{p+1}}\tilde{Y}_{T}^{\mu_{1}...\mu_{p+1}}  -\frac{1}{2p!}m_{X}^{2}\tilde{X}_{T\mu_{2}...\mu_{p+1}}\tilde{X}_{T}^{\mu_{2}...\mu_{p+1}}\right]+ \nonumber 
\\&&+ \int \e^{\alpha_{p}A}u(z)^{2}dz\int d^{D-1}x\left[-\frac{1}{2p!} \tilde{Y}^{\mu_{1}...\mu_{p}}\tilde{Y}_{\mu_{1}...\mu_{p}} -\frac{1}{2(p-1)!} m_{p-1}^{2}\tilde{X}_{\mu_{2}...\mu_{p}}\tilde{X}^{\mu_{2}...\mu_{p}}\right]+ \nonumber
\\&&+ \int \e^{\alpha_{p}A}F^{2}dz\int d^{D-1}x\left[-\frac{1}{2(p+1)!}\tilde{Y}_{L\mu_{1}...\mu_{p+1}}\tilde{Y}_{L}^{\mu_{1}...\mu_{p+1}} -\frac{1}{2p!}m_{L}^{2}\tilde{X}_{L\mu_{1}...\mu_{p}}\tilde{X}_{L}^{\mu_{1}...\mu_{p}}\right]\nonumber
\\&&+\frac{1}{2(p-1)!}\int \e^{\alpha_{p}A} \left(\e^{-(D-2p)A}(\e^{(D-2p)A}u(z))'\right)'u(z)dz\int d^{D-1}x\tilde{X}_{\mu_{2}...\mu_{p}5}\tilde{X}^{\mu_{2}...\mu_{p}5}, \label{Arp} 
\end{eqnarray}
\end{small}
where we have defined $m_{L}$ by 
\begin{equation}
\e^{-\alpha_{p}A}\left(\e^{\alpha_{p}A}F'\right)' -(M^{2}+c\delta(z))\e^{2A}F(z) = -m_{L}^{2}F(z).
\end{equation}
The results obtained in this section shows that the above action reduces to action of transversal massless $p$-form
\begin{eqnarray}
S_{D-1} &=& \int d^{D-1}x\left[-\frac{1}{2(p+1)!}\tilde{Y}_{T\mu_{1}...\mu_{p+1}}\tilde{Y}_{T}^{\mu_{1}...\mu_{p+1}} \right]
\end{eqnarray}
on the brane if $\nu>1$ and $D \neq 2p+1$. This case in similar to 1-form case, where only the transverse part is located, i.e., the effective range has already fixed the Lorentz gauge.
The five dimensional action reduces to the standard action of massless $p$-form
\begin{eqnarray}
S_{D-1} &=& \int d^{D-1}x\left[-\frac{1}{2(p+1)!}\tilde{Y}_{\mu_{1}...\mu_{p+1}}\tilde{Y}^{\mu_{1}...\mu_{p+1}}\right]
\end{eqnarray}
on the brane if $1<\nu\leq2$ and $D = 2p+1$;
and to the action of massless $p$-form with a massless $(p-1)$-form 
\begin{eqnarray}
S_{D-1} &=& \int d^{D-1}x\left[-\frac{1}{2(p+1)!}\tilde{Y}_{\mu_{1}...\mu_{p+1}}\tilde{Y}^{\mu_{1}...\mu_{p+1}} -\frac{1}{2p!} \tilde{Y}^{\mu_{1}...\mu_{p}}\tilde{Y}_{\mu_{1}...\mu_{p}}\right]
\end{eqnarray}
on the brane if $\nu >2$ and $D = 2p+1$, where $\tilde{Y}_{\mu_{1}...\mu_{p+1}}$ is the propagator of the field $ \tilde{X}_{\mu_{1}...\mu_{p}} \equiv \tilde{X}_{T\mu_{1}...\mu_{p}} +\tilde{X}_{L\mu_{1}...\mu_{p}}$.
This case is similar to Kalb-Ramond case, where the localized $p$-form in composed of booth parts of 5-dimensional field, thus there is no gauge fixing.

\section{The $p$-form Case in a Smooth Warp Factor Scenario}
\begin{figure}[h!]
 \centering
 \includegraphics[scale=0.4]{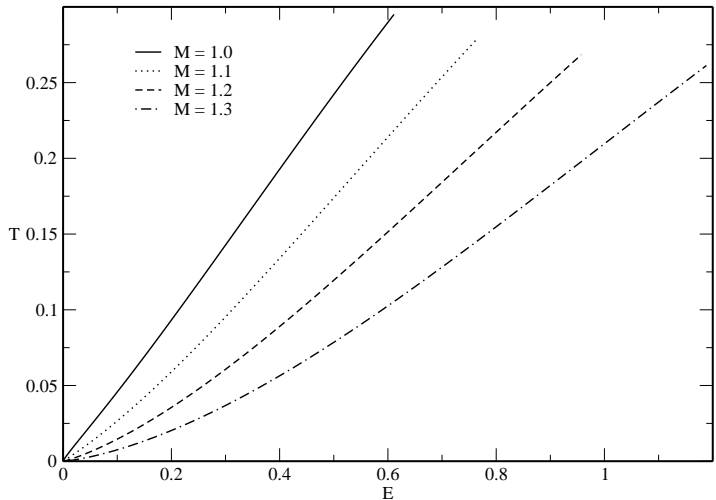}
 % Tsmooth-vector.pdf: 792x612 pixel, 72dpi, 27.94x21.59 cm, bb=0 0 792 612
 \caption{Plot of transmission coefficient for 1-form field in 5-dimension as function of energy, $E = m^{2}_{X}$, for different values of mass parameter $M$. We have fixed $ k = \beta = 1$.}
 \label{fig: tsmoothvector}
\end{figure}

\begin{figure}[h!]
 \centering
 \includegraphics[scale=0.4]{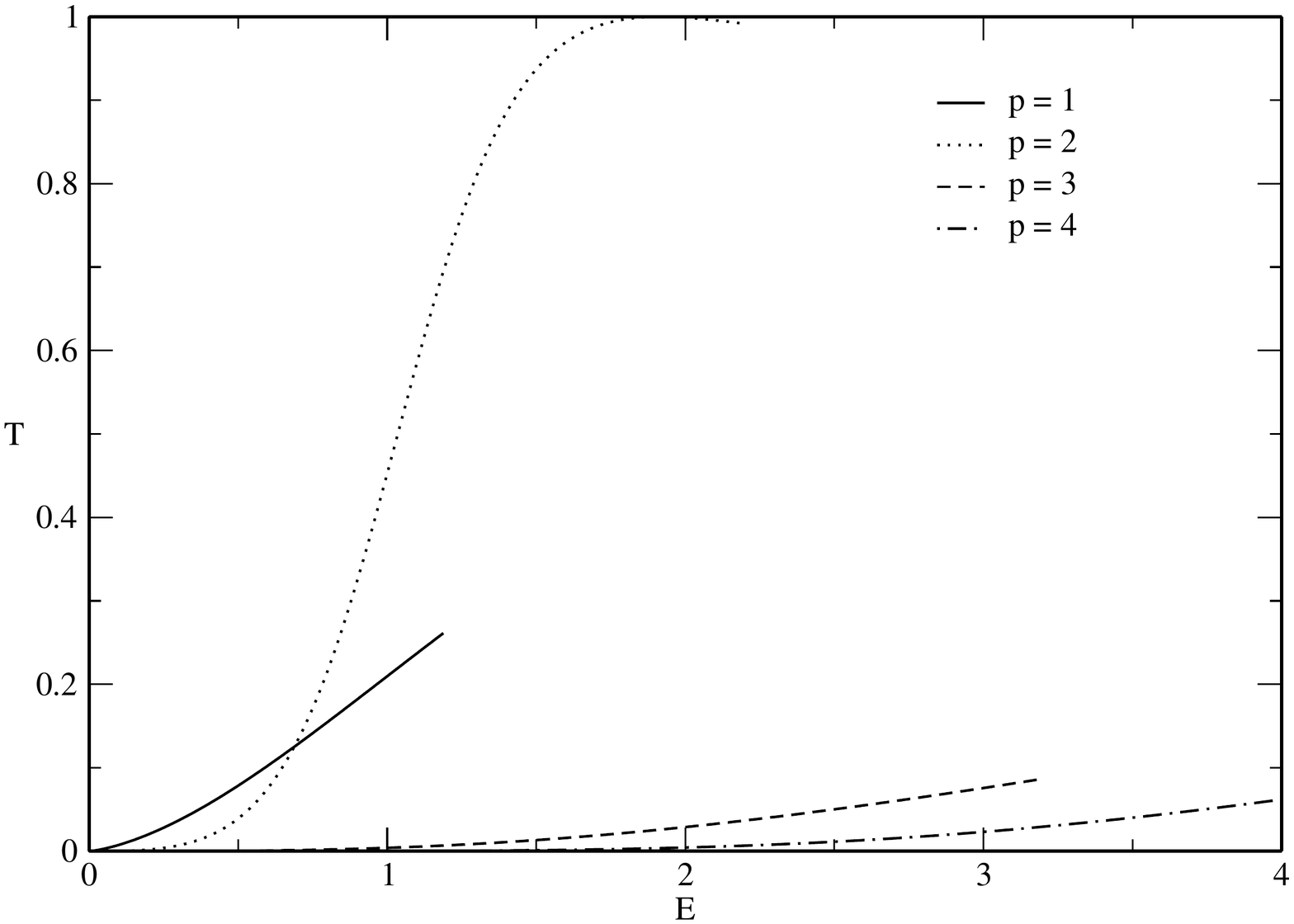}
 % Tsmooth-p.pdf: 792x612 pixel, 72dpi, 27.94x21.59 cm, bb=0 0 792 612
 \caption{Plot of transmission coefficient as function of energy, $E = m^{2}_{X}$, for different $p$-forms in five dimensions. We have fixed $ k = \beta = 1$ and $M = 1.3$.}
 \label{fig: tsmoothp}
\end{figure}

In this section we investigate the localization of a $p$-form field in a smooth warp factor scenario. Since the metric can be written in a conformal form we can use all results obtained in
previous section which do not use the explicit form of the warp factor. In this section we will use the following smooth warp factor \cite{Du:2013bx, Melfo:2002wd}
\begin{equation}
  A(z) = -\frac{1}{2}\ln\left[\left(kz\right)^{2}+\beta\right],
\end{equation}
which recover the Randall-Sundrum metric at large $z$. Using this metric in eq. (\ref{potp}) we obtain the Schrödinger's potential for transversal part of $p$-form
 \begin{equation}\label{potpsmo}
U(z)=\left(\frac{\alpha_{p}^{2}}{4} +\alpha_{p}\right)\frac{\left(kz\right)^{2} k^{2}}{[\left(kz\right)^{2}+\beta]^{2}}-\frac{\alpha_{p}}{2}\frac{k^{2}}{[\left(kz\right)^{2}+\beta]}+\frac{M^{2}}{\left[(kz)^{2}+\beta\right]} +\frac{c}{\beta^{2}}\delta(z), 
\end{equation}
For massless mode of transversal $p$-form the Schrödinger's equation with above potential provides the following convergent solution
 \begin{equation}\label{solfpsmo}
 \psi = f_{0}\left[(kz)^{2} +\beta\right]^{(1-2\nu)/4}\hip{\frac{-1+\nu}{2}, \frac{2+\nu}{2};1+\nu; \frac{\beta}{(kz)^{2}+\beta}},
 \end{equation}
where $f_{0}$ is a constant and $\nu$ is the same of (\ref{nup}). To satisfy the boundary conditions at origin is necessary fix the four dimensional mass parameter, $\mu^{2} = c/\sqrt{\beta}$, 
\begin{equation}\label{csmooth}
 \mu^{2} = -2k\frac{(\nu^{2}-1)}{\nu}
\end{equation}
The solution of massive modes of  transversal $p$-form can not be found analytically. To obtain information about this state we use the transference matrix method to evaluate the transmission coefficient. 
The behavior is illustrated in fig. \ref{fig: tsmoothvector} for 1-form with some values of mass parameter and in fig. \ref{fig: tsmoothp} for different $p$-forms. Both figures do not exhibit  peaks, indicating the no existence of unstable modes.  

For massless mode of reduced $(p-1)$-form the eq. (\ref{potp-1}) provide the following potential
 \begin{equation}\label{potp-1smo}
U(z)=\left(\frac{\alpha_{p-1}^{2}}{4} +\alpha_{p-1}\right)\frac{\left(kz\right)^{2} k^{2}}{[\left(kz\right)^{2}+\beta]^{2}}-\frac{\alpha_{p-1}}{2}\frac{k^{2}}{[\left(kz\right)^{2}+\beta]}+\frac{M^{2}}{\left[(kz)^{2}+\beta\right]} +\frac{c}{\beta^{2}}\delta(z), 
\end{equation}
which give us the same solution (\ref{solfpsmo}), with new multiplicative constant $\bar{f}_{0}$. Due the warp factor be smooth the boundary condition at $z=0$ is the same of $p$-form case, i.e., the same fixation  (\ref{csmooth}) provide
a localized solution massless mode of reduced $(p-1)$-form if $\nu >2$.

The longitudinal part of $p$-form is determined by eq. (\ref{tracelesspform}). Imposing the separation of variables in the form  $X^{\mu_{1}...\mu_{p}}_{L}(z,x) = F(z)\tilde{X}^{\mu_{1}...\mu_{p}}_{L}(x)$ we
obtain
\begin{eqnarray}
F(z) &=& F_{0}k z \left[(kz)^{2}+\beta\right]^{(\alpha_{p}-2\nu-5)/4} \times \nonumber
\\&&\times\left[(\alpha_{p}+2\nu+1)\left[(kz)^{2}+\beta\right] \, _2F_1\left(\frac{\nu-1}{2},\frac{\nu+2}{2};\nu+1;\frac{\beta}{(kz)^{2} +\beta}\right)+\right. \nonumber
\\&&+\left.\frac{\beta (\nu-1) (\nu+2)}{\nu+1} \, _2F_1\left(\frac{\nu+1}{2},\frac{\nu+4}{2};\nu+2;\frac{\beta}{(kz)^{2}+\beta}\right)\right]
\end{eqnarray}
where $F_{0}$ is a constant proportional to $\bar{f}_{0}$. Due the $(p-1)$-form is localized if $\nu>2$, the longitudinal part of $p$-form will  be localized if $\nu>1$.

The results obtained in this section shows that the action (\ref{Arp}) reduces to 
\begin{eqnarray}
S_{D-1} &=& \int d^{D-1}x\left[-\frac{1}{2(p+1)!}\tilde{Y}_{\mu_{1}...\mu_{p+1}}\tilde{Y}^{\mu_{1}...\mu_{p+1}}\right]
\end{eqnarray}
on the brane if $1<\nu\leq2$;  and to
\begin{eqnarray}
S_{D-1} &=& \int d^{D-1}x\left[-\frac{1}{2(p+1)!}\tilde{Y}_{\mu_{1}...\mu_{p+1}}\tilde{Y}^{\mu_{1}...\mu_{p+1}} -\frac{1}{2p!}\tilde{Y}_{\mu_{1}...\mu_{p}}\tilde{Y}^{\mu_{1}...\mu_{p}}\right]
\end{eqnarray}
on the brane if $\nu>2$, where $\tilde{Y}_{\mu_{1}...\mu_{p+1}}$ is the propagator of the $p$-form $ \tilde{X}_{\mu_{1}...\mu_{p}} \equiv \tilde{X}_{L\mu_{1}...\mu_{p}} +\tilde{X}_{L\mu_{1}...\mu_{p}}$.
Due the warp factor is smooth its does not change the boundary condition at the origin, so that both parts of the field in five dimensions and the $(p-1)$-form are localized with the same fixation (\ref{csmooth}).
Thus the localized $p$-form contains booth parts of 5-dimensional field, i.e., there is no gauge fixing.

\section{Conclusion}

In this paper we further developed the model proposed by  Ghoroku and Nakamura in \cite{Ghoroku:2001zu} and apply it to $p-$form fields. In  Sec. II we show that the definitions of the transverse and longitudinal parts are sufficient to decouple the equations of motion, being unnecessary to impose the parity used in previous works. This simplifies the assumptions needed by the model. We calculate the value of the coupling parameter with the brane, $c$, which localizes the zero mode of the transversal part of the vector field.  We show that the massive modes of the transverse part of the vector field is non-localized, and calculating the transmission coefficient we show that there are no massive unstable modes. This is a very unexpected property of this model that also happens to the other cases presented here.  It was also shown that  the longitudinal part of the vector field and the scalar field are non-localized. By a complex transformation in massive solutions we conclude that there are no localized 
tachyonic modes. These results are interesting because we find that only the gauge field is observed in four dimensions.

Using the same procedure for the Kalb-Ramond field we find the value of the coupling constant with the brane that localizes the zero mode of the transversal part of the field. We showed that unlike the previous case the same coupling constant localizes the zero modes of the transversal part of the Kalb-Ramond field and the vector field. This is an interesting possibility since that allows to construct other kinds of models in five dimensions. For example, the KR field can be thought as a source of torsion in the four dimensional model,with a localized vector field.  Yet in Sec. III we show that no massive mode is localized, as well as tachyonic modes. An analysis of transmission coefficient indicates that are no unstable massive modes.

In Sec. IV we generalized the procedure used in Sec. III to the $p$-form case in a $D$-dimensional bulk. We compute the value of the coupling parameter with the brane, $c$, which provides a localized solution to massless mode of transversal part of $p$-form. We show that, for all $D$ and $p$, all massive and tachyonic modes are non-localized. An analysis of transmission coefficient indicates that there are no unstable massive modes. Also in this section we found a relation between $D$ and $p$ which localize the reduced $(p-1)$-form and massless mode of longitudinal part of $p$-form, in agreement with results obtained previously.  In this scenario we show that if  $D>2p+3$ or $D< 2p-1$ the zero mode  of transversal part of $p$-form are localized without five dimensional massive term, i.e., with gauge invariance in the bulk.

Finally, in Sec. V, we use the procedure used in Sec. IV for a smooth warp factor. Due to the fact that the metric is smooth we show that the same fixation of coupling parameter, $c$, localize all massless modes, differently of Randall-Sundrum case. This result showed that even in smooth brane scenario the coupling constant is necessary to localize the zero modes, but now it can not be understood as a coupling with the brane. We found analytical solution to the massless case but we are not able to find the same to the massive modes. To overcome this problem we used the method of the transfer matrix to obtain the transmission coefficient. An analysis of the transmission coefficient leads to the conclusion that there is no unstable massive modes.

It should be pointed out that the results obtained in this manuscript opens possibilities for new directions. First of all it is important to understand the origin of the 
boundary terms. In this directions some of us recently proposed a gravitational origin \cite{Alencar:2014moa, Alencar:2014fga, Jardim:2014xla}. Despite of this the proposed model also has 
parameters to be fine-tuned what should be understood from a more fundamental view point. Anyway this kind of coupling seems to point to something interesting since
it also works for other kinds of fields, such as Elko spinors. In fact this suggest that non-minimal couplings play a central role in RS scenarios \cite{Alencar:2015awa}. Another possibility is
considering models with interactions. This can raise new important question such unitarity and stability. This kind of problem has been studied by 't Hooft in Ref. 
\cite{'tHooft:1971rn}. All of this is out of the scope of the present paper but are under investigation by us.

\section*{Acknowledgments}

We acknowledge the financial support provided by Funda\c c\~ao Cearense de Apoio ao Desenvolvimento Cient\'\i fico e Tecnol\'ogico (FUNCAP), the Conselho Nacional de 
Desenvolvimento Cient\'\i fico e Tecnol\'ogico (CNPq) and FUNCAP/CNPq/PRONEX. The author acknowledge the referees for suggestions and for discussion in important points.

\end{document}